\documentclass[a4paper,11pt]{article}
\usepackage{jinstpub} 
\usepackage{lineno}

\title{Characterisation of SiC radiation detector technologies with synchrotron X-rays}

\author[a,1]{I. Lopez Paz,\note{Corresponding author.}}
\author[a]{C. Fleta,}
\author[a]{J. M. Raf\'i,}
\author[a]{G. Rius,}
\author[a]{P. Godignon,}
\author[a]{G. Pellegrini,}
\author[a]{S. Mena,}
\author[a]{M. Jimenez,}
\author[a]{A. Henao,}
\author[a]{J. Bravo,}
\author[b]{R. Boer,}
\author[b]{B. Molas}
\author[a]{and C. Guardiola}

\affiliation[a]{Institute of Microelectronics of Barcelona, IMB-CNM (CSIC),\\Carrer dels Til·lers s/n, 08193 Cerdanyola del Vall\`es (Bellaterra), Barcelona, Spain}
\affiliation[b]{ALBA Synchrotron, Carrer de la Llum 2-26, 08290 Cerdanyola del Valles, Spain}
\emailAdd{ivan.lopez@imb-cnm.csic.es}

\abstract{
To cope with environments with high levels of radiation, non-silicon semiconductors such as silicon carbide detectors are being proposed for instrumentation.

4H-SiC diodes for radiation detection have been fabricated in the IMB-CNM Clean Room, for which different strategies to define the electrical contact of the implants had been implemented, in an attempt to optimise the technology for, e.g., medical applications or low energy radiation detection, as the material choice can affect the sensitivity of the device. Among these technologies, it is included an epitaxially-grown graphene layer as part of the electrical contact.

In this paper, a selection of four configurations of the IMB-CNM SiC diodes are characterised in terms of radiation detector response.
Photodiode performance under 20~keV X-rays irradiation in the XALOC beam line at ALBA Synchrotron is presented. Over-responses in the range of 12-19\% linked to the interaction of the radiation with the metallic layers are observed. A good uniformity response as well as a good linearity at 0~V bias is reported, even in the under-depleted devices.
This work exemplifies the good performance of SiC detectors fabricated at IMB-CNM specifically for low-energy X ray characterization at high X-ray intensities.
}


\begin{document}
\maketitle
\flushbottom

\section{Introduction}
\label{sec:intro}

Over the last years the use of Silicon Carbide as a material for detector fabrication has increased as an alternative to Silicon, owing to its higher thermal conductivity, lower leakage current and larger displacement energy threshold~\cite{SiC-review}. 
For this reason, the use of this wide band-gap semiconductor has been studied for applications such as High Energy Physics~\cite{SiC-rad-det,SiC-irrad} as well as nuclear applications~\cite{SiC-neutron,SiC-nuclear,Prez2024}, synchrotron beam monitoring~\cite{SiC-Synchrotron} and medical dosimetry~\cite{SiC-Graphene_Med, SiC-UHDR, Milluzzo2024}.
At the Institute of Microelectronics of Barcelona (IMB-CNM), the SiC radiation detector technology is under investigation~\cite{SiC-irrad,SiC-nuclear,otero2022}. 

Firstly, IMB-CNM has developed 4-quadrant SiC photodiodes for beam position monitoring in collaboration with the ALBA and ESRF synchrotrons~\cite{SiC-4q, SiC-irrad}, showing an improved performance at variable temperatures and visible light illumination conditions and superior radiation hardness to similar Si photodiodes.

Secondly, SiC photodiodes like the ones in \cite{SiC-irrad} were irradiated with an X-ray tube at 50~kV with the detectors at 0~V bias and in a 9~MeV electron beam from a Varian Clinac 2100 accelerator in a first study. The response of the SiC diodes to radiation showed excellent stability and reproducibility. The linearity deviation with the integrated dose delivered was less than 1.5\%, while the medium-term stability with integrated dose up to 18~kGy using X-rays was better than 0.3\% in pristine (i.e., non pre-irradiated) diodes, which is much better than for equivalent Si-based photodiodes.

Thirdly, in the framework of the EMPIR EU UHDPulse project~\cite{EMPIR}, IMB-CNM has already designed and manufactured SiC detectors that have been tested successfully under FLASH conditions in an electron beam at PTB (Germany). A tested SiC detector showed a good linearity of collected charge per pulse (1.5-3~$\mu$s) in a 20~MeV electron beam~\cite{SiC-UHDR}.

In parallel, other groups are exploring the fabrication of SiC diodes for high temperature applications~\cite{HighT-SiC} and novel geometries, such as SiC membranes for synchrotron beam position monitoring~\cite{Sensic1, Sensic2}.

Many radiation detector prototypes have been fabricated~\cite{SiC-Graphene_TCT, SiC-UHDR, Gaggl2023}, exploring different configurations, such as combinations of metal contacts, or graphene as an interface layer with the active volume, to accommodate a variety of applications.
Application-oriented considerations include removing all metallic structures in medical dosimeters, which means that the contribution of secondary interactions in X-ray therapies also reduce, thus avoiding false contributions and distortions of the main signal~\cite{Metal-dose-enhancement}. The presence of metals can also prevent the possibility to detect UV light~\cite{SiC-Graphene_TCT}.
Moreover, different epitaxy layer thickness may be required for each purpose: a thick active region allows for higher signals from highly penetrating particles, while shallow epitaxies are favorable for un-biased current mode measurements, as the built-in voltage from the junction is enough to deplete the full volume.

In this study, a selection of the diodes produced at IMB-CNM - with and without metal, with graphene contacts and either 3 or 50~$\mu$m epitaxy - are characterised in the ALBA Synchrotron~\cite{ALBA_1,ALBA_2}, thus profiting from the adjustable mono-energetic X-ray beam and small beam size, as its radiation source characteristics makes the response of the hereby proposed instrumentation sensitive to the detector design and material choice.

\section{Silicon Carbide Diodes}
\label{sec:duts}

\begin{figure}[htbp]
\centering
\includegraphics[width=1\textwidth]{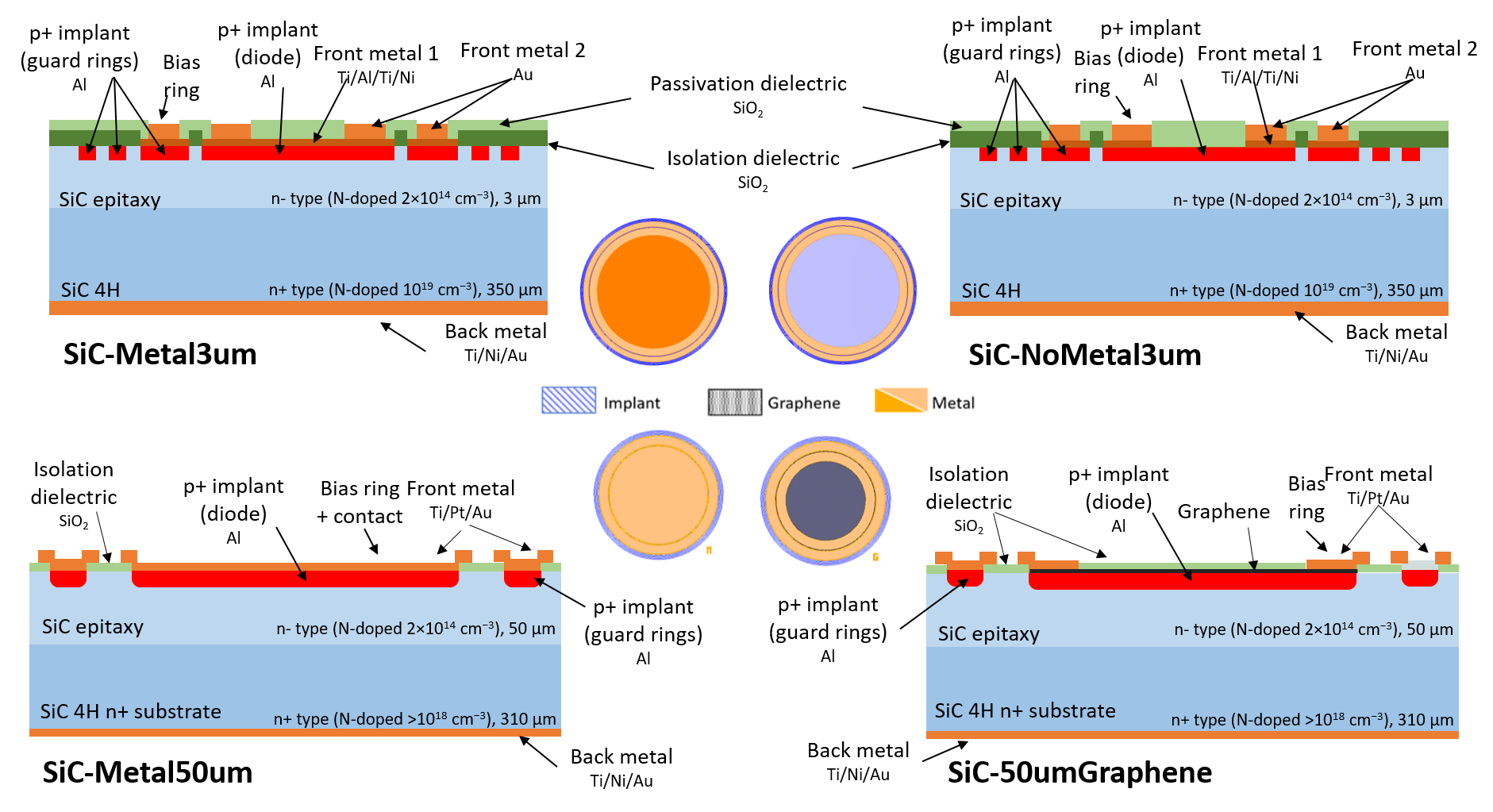}
\caption{Cross section (corners) and top view (centre) schematics of the DUTs. Sketches adapted from~\cite{SiC-Graphene_Med} and \cite{SiC-UHDR}.\label{fig:dut-schematic}}
\end{figure}

Silicon Carbide (4H-SiC politype) commercial epitaxial wafers with 3 and 50~$\mu$m epitaxy were used to the fabricate diodes. In all instances, aluminum ions were implanted in-house to produce the pn-junction.
Different strategies for the electrical contact with the diode implant have been utilised in the devices under test (DUTs), as shown in Figure~\ref{fig:dut-schematic}.

In the 3~$\mu$m epitaxial wafer, a metal stack is utilised forming a 2.2~mm diameter diode (SiC-Metal3um, Fig.~\ref{fig:dut-schematic}, top-left)~\cite{SiC-UHDR}. The circular metal layer is covered by a passivation dielectric layer except for a bias ring to allow for wirebonding. Alongside, a diode was fabricated featuring no metal over the active area, except for the bias ring (SiC-NoMetal3um, Fig.~\ref{fig:dut-schematic}, top-right), which was used in some of the experiments featured in this document.

The wafer with 50~$\mu$m epitaxy feature either an epitaxialy-grown graphene layer~\cite{SiC-Grphn_1,SiC-Grphn_2} or a metal stack (SiC-Graphene50um and SiC-Metal50um respectively,  Fig.~\ref{fig:dut-schematic}, bottom), defining both a 1~mm diameter active area~\cite{SiC-Graphene_Med}. The sample containing a graphene layer features a metallic ring with an inner diametre of 800~$\mu$m to allow electrical contact with wirebonding, similar to the SiC-Metal3um case, while the implantation on SiC-Metal50um is fully covered by the same Ti/Pt/Au stack.
All metals and thicknesses for each DUT are reported in Table~\ref{tab:duts}.
The back side, in all cases, is fully metallised with a combination of Ti/Ni/Au.

\begin{table}[htbp]
\centering
\caption{List of DUTs for the experiments.\label{tab:duts}}
\smallskip
\begin{tabular}{l|cccc}
\hline\hline
Sample Name& Active layer& Contact material& Bias ring &Ref.\\
\hline
SiC-Graphene50um& 50~$\mu$m-epi SiC    & Graphene  &  Ti/Pt/Au  & \cite{SiC-Graphene_Med}\\
SiC-Metal50um   & 50~$\mu$m-epi SiC& Ti/Pt/Au&Ti/Pt/Au& \cite{SiC-Graphene_Med}\\
SiC-Metal3um    & 3~$\mu$m-epi SiC & Ti/Al/Ti/Ni        & Ti/Au     & \cite{SiC-UHDR}\\
SiC-NoMetal3um    & 3~$\mu$m-epi SiC & Ti/Al/Ti/Ni & Ti/Au     & \cite{SiC-UHDR}\\
                &               & (only under bias ring) &   &   \\\hline
Si-Reference    & \multicolumn{3}{c}{Hamamatsu Silicon photodiode (S10043) } &\cite{Hamamatsu-Si}\\
\hline \hline
\end{tabular}
\end{table}

Each diode was glued to a PCB with silver paint to allow grounding to the back of the diodes. Moreover, SiC-Graphene50um and SiC-Metal50um were fully covered with epoxy to remove possible surface currents, while for sample SiC-Metal3um and SiC-NoMetal3um only the wirebonds were protected with epoxy. Included in the experiment was a commercial Hamamatsu silicon 10$\times$10~mm$^2$ photodiode (S10043)~\cite{Hamamatsu-Si} was utilised for reference (Si-Reference), as its response was calibrated with the beam power at the BESSY II facility.

\section{Irradiation Set-up}
\label{sec:setup}
The devices were tested in the BL13-XALOC~\cite{XALOC} beamline at the ALBA Synchrotron~\cite{ALBA_1,ALBA_2}. The ALBA Synchrotron accelerates electrons up to 3~GeV with a nominal current of 250~mA to generate photon beams in each experimental area. 
The beam current in the accelerator ring decays over time but is controlled with periodic injections to maintain a constant intensity.

\begin{figure}[htbp]
\centering
\includegraphics[width=.6\textwidth]{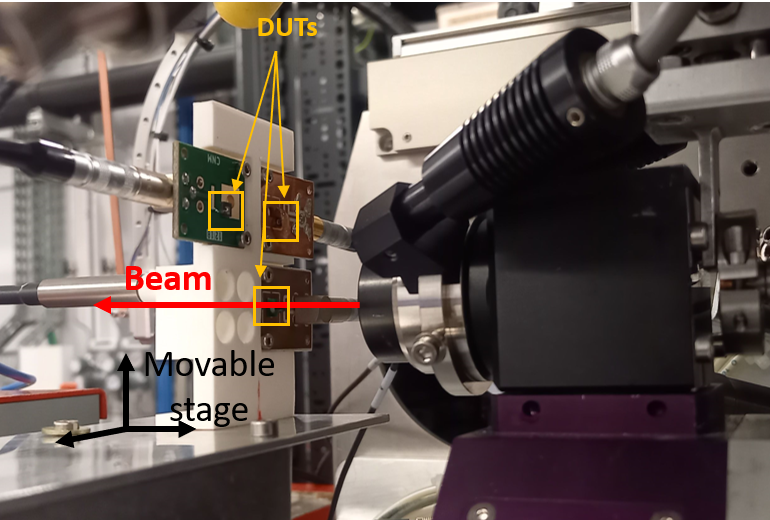}
\caption{Experimental set-up at BL13 XALOC beam line. \label{fig:setup}}
\end{figure}

The XALOC beamline (BL13) provides X-ray beams of energies ranging from 4.6 to 23~keV - unless stated otherwise, 20~keV is used hereafter-, and beam spot sizes from 50$\times$10 to 300$\times$300~$\mu$m$^2$. The beam was focused to a 50$\times$50~$\mu$m$^2$ size for our tests. The intensity can be regulated by means of a set of attenuators~\cite{XALOC}, varying the beam current reaching the DUTs (transmission), set to 100\% unless stated otherwise. The DUTs were mounted in a micrometer-precision 3-axis motorised stage as shown in Figure~\ref{fig:setup}.

The readout of the detectors is performed via a custom-made ALBA Em electrometer, with four current inputs with a manually configurable current range of 100~pA to 1~mA and a 1~kSamples per second resolution.
Each measurement point is the average current recorded by the electrometer over 200~ms. Thus, the detector response is faster than the temporal resolution of the electrometer~\cite{SiC-UHDR}.
At the end of the experiment no indication of radiation damage was observed.

All measurements have been taken with a bias voltage of 0~V. This is due to a limitation on the electrometer, which does not allow for high voltage biasing. Nevertheless, the built-in voltage of the detectors is enough to fully deplete the 3~$\mu$m active depth of the devices SiC-Metal3um and SiC-NoMetal3um, and the devices SiC-Metal50um and SiC-Graphene50um already have shown a good photocurrent response while not fully depleted~\cite{SiC-Graphene_Med}.

\section{SiC Device Performance}
\label{sec:sicresults}

The photocurrent, stage positions, beam energy and beam transmission are monitored while scanning either parameter. For the data treatment, the dark current is subtracted from the detector response, as measured with the highest precision electrometer setting, in the order of 10-30~pA.
Then, the readings are scaled with the machine beam current to the average value of 251.44~mA as it is continuously monitored, to correct for the variation of intensity due to injection and decay in the synchrotron accelerator ring. This beam current has a maximum deviation over time of $\pm$1.93~mA, which only represents a 0.7\%.

\subsection{Position Response}
\label{sec:position}
\begin{figure}[htbp]
\centering
\includegraphics[width=.6\textwidth]{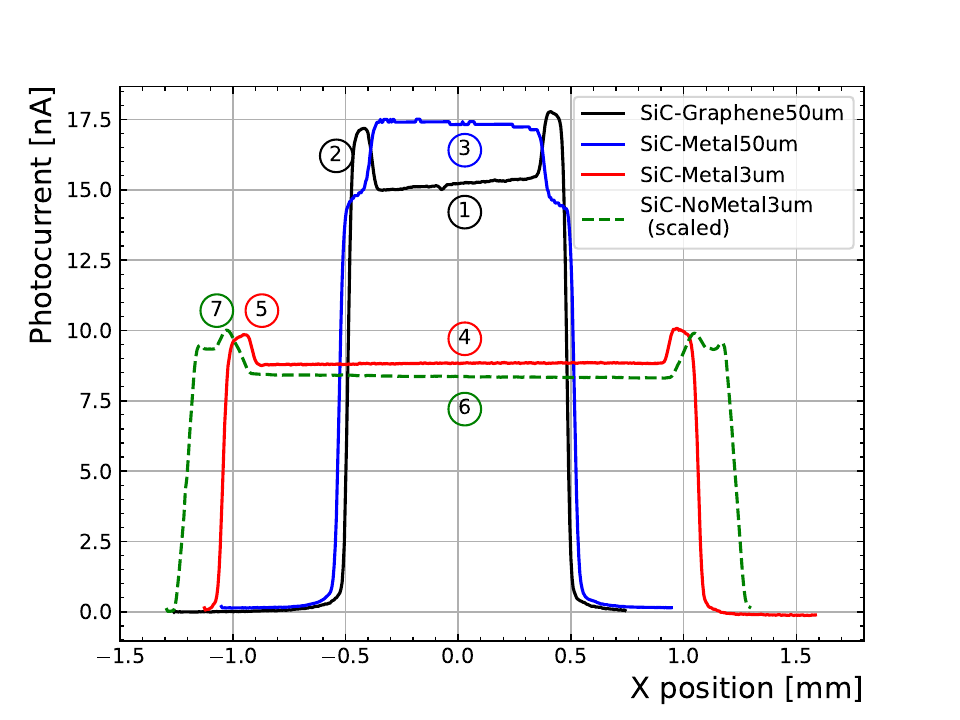}
\caption{Photocurrent response as a function of beam position across the studied SiC diodes. Note that the widths do not exactly correspond to the respective diameters, as the measurements were not performed along an exact diameter. The regions with different material composition are marked with numbers with the same colour code as the data for reference in the text. The SiC-NoMetal3um was scaled (dashed line) to match the beam intensity of the other measurements (see text).\label{fig:linescan}}
\end{figure}

Figure~\ref{fig:linescan} shows the response of the DUTs as a function of beam position.
The generated photocurrent is monitored while the detectors are moved across the horizontal direction with respect to the 50$\times$50~$\mu$m$^2$ X-ray beam, covering the full active area of the devices. 

The SiC-NoMetal3um was measured in a different beam campaign, with the same beam energy and size, but different intensity. To correct for the intensity mismatch observed across days the measurement data is normalised. The current at the SiC-Metal3um was measured in the inner region at the same conditions as the SiC-NoMetal3um. Then, together with the SiC-Metal3um data from the first campaign, the measurement from sample SiC-NoMetal3um was scaled.
The bias ring at $x=\pm$1~mm (labelled position 7 in Fig.~\ref{fig:linescan}) in the SiC-NoMetal3um has the same metal combination as the bias ring in SiC-Metal3um (labelled position 5) and show the same response after scaling, showing the validity of the normalisation.

The higher sensitivity of the 50~$\mu$m epitaxial diodes (positions 3 and 1) with respect to their 3~$\mu$m counterparts (4 and 6) is expected due to a difference in active volume at 0~V. While the latter is fully depleted and thus collects charges from the 3~$\mu$m of epitaxy, approximately $\sim$6~$\mu$m~\cite{SiC-nuclear} (estimated from the capacitance) are depleted via the built-in voltage from the junction in the 50~$\mu$m thick epitaxy devices.

Additionally, the higher central response about 15\% higher of the SiC-Metal50um (3) device with respect to SiC-Graphene50um (1) can be explained by the presence of high-$Z$ metals such as Au and Pt in the detector surface. The cross section for the photoelectric effect absorption grows approximately as $Z^4$, which therefore generates secondary low energy electrons which are completely absorbed in the active volume.
This is confirmed by the over-response observed in the SiC-Graphene50um in the ring region (2), which shows the same metal combination as the SiC-Metal50um over the full active area, yielding a similar photocurrent. 

A similar effect of 12-14\% and 18-19\% is seen in the SiC-Metal3um (5) and the SiC-NoMetal3um (7) respectively, at the region corresponding to their respective bias ring as they also contain a 100~nm gold layer, where a higher current is observed. The difference across these two devices in the central region (4 and 6) are at the same time attributed to the presence of metal or lack thereof. 
The broader outer over-response region on the SiC-NoMetal3um (7) than that of the SiC-Metal3um (5) is a consequence of the interconnection between the bias ring and the guard-ring surrounding the active region of the same metal composition, effectively increasing the sensitive area.
The SiC-Metal50um on the other hand showcases lower sensitivity near the edge of the metallic contact than in its central region. This is still under investigation, as its design contains no feature corresponding to that area.

The uniform response of the SiC-NoMetal3um (6) even when the current collection is done across the implantation, removes the need of metal, which is relevant for e.g. medical dosimetry, as it improves on its equivalency with tissue. However, practically a lack of metal may affect its performance after irradiation, as there are indications of loss of dopant concentration~\cite{SiC-irrad}. In addition, missing a conductive layer has been shown to be detrimental for pulse integrity~\cite{SiC-Graphene_TCT}.


The slight slope observed in the response in both SiC-Metal50um (3) and SiC-Graphene50um (1) are due to the presence of uneven epoxy thickness over their active area, absorbing varying amounts of radiation. Other than that, both devices show a constant current across their active region, as expected from 369~nm UV laser~\cite{SiC-Graphene_TCT} and 3.5~MeV He$^+$ IBIC~\cite{SiC-nuclear} measurements in similar diodes for the graphene and metal samples respectively. The remaining detectors, the SiC-Metal3um and the SiC-NoMetal3um as they are not covered by epoxy, show a uniform response across their central region (4 and 6).


\subsection{Linearity Evaluation}
\label{sec:currentscan}

\begin{figure}[htbp]
\centering
\includegraphics[width=.6\textwidth]{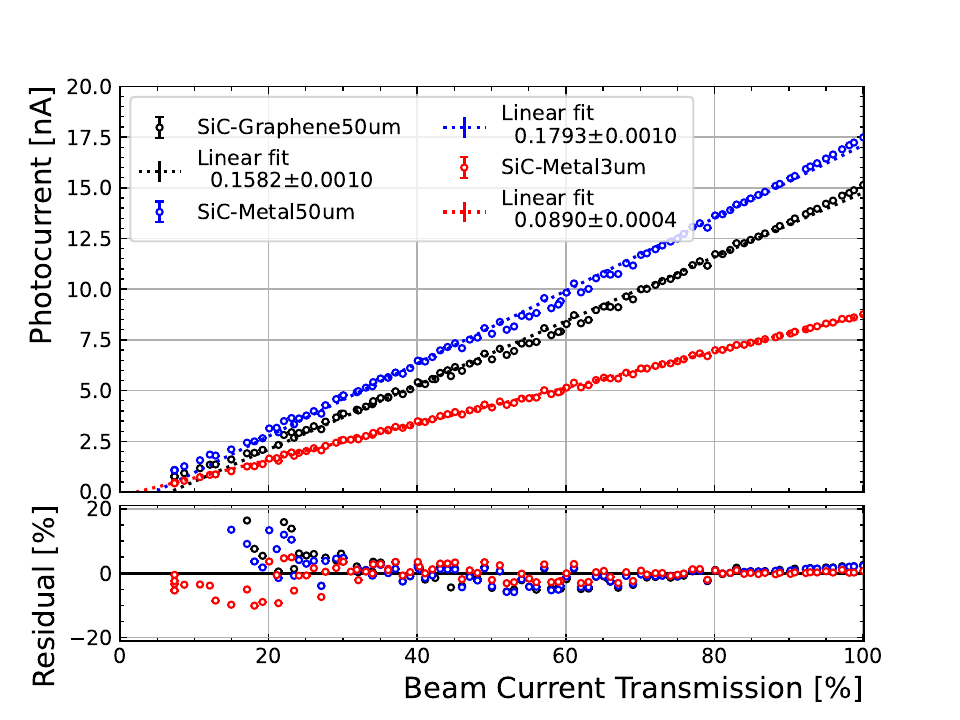}
\caption{Response as a function of transmission across the studied SiC diodes, fitted linearly. The residuals are showed below. \label{fig:currentscan}}
\end{figure}

By means of a set of 12 attenuators~\cite{XALOC}, the beam intensity reaching the DUTs can be controlled in a range of 7.36 to 100\% of the maximum intensity. 
The response of the diodes, with the beam focused at their central area, are correlated to the beam intensity (see Figure~\ref{fig:currentscan}). A straight line is fitted on each of the data sets, and their respective residuals (i.e. the relative difference between fit and data) are shown to assess the linearity of their response.

While the residuals observed with the SiC-Metal3um remain better than 10\% over the full beam current range, the 50~$\mu$m-epitaxy detectors diverge from the fit at the lowest intensities (i.e. transmission <20\%), possibly a consequence of under-depletion or non-considered uncertainties in the beam intensity. 
The deviations from linearity at higher intensities are consistent across detectors, which indicate a small (5\%) discrepancy between the transmissions reported by the facility and the beam intensity.
On the other hand, the detectors have performed at percent-level linearity in other experimental settings~\cite{SiC-Graphene_Med, SiC-UHDR}, which indicates that these deviations may also be related to the precision limit of the X-ray beam attenuation.
Nevertheless, no indication of saturation is observed in tested conditions.

\subsection{Energy Response}
\label{sec:energyscan}

\begin{figure}[htbp]
\centering
\includegraphics[width=.7\textwidth]{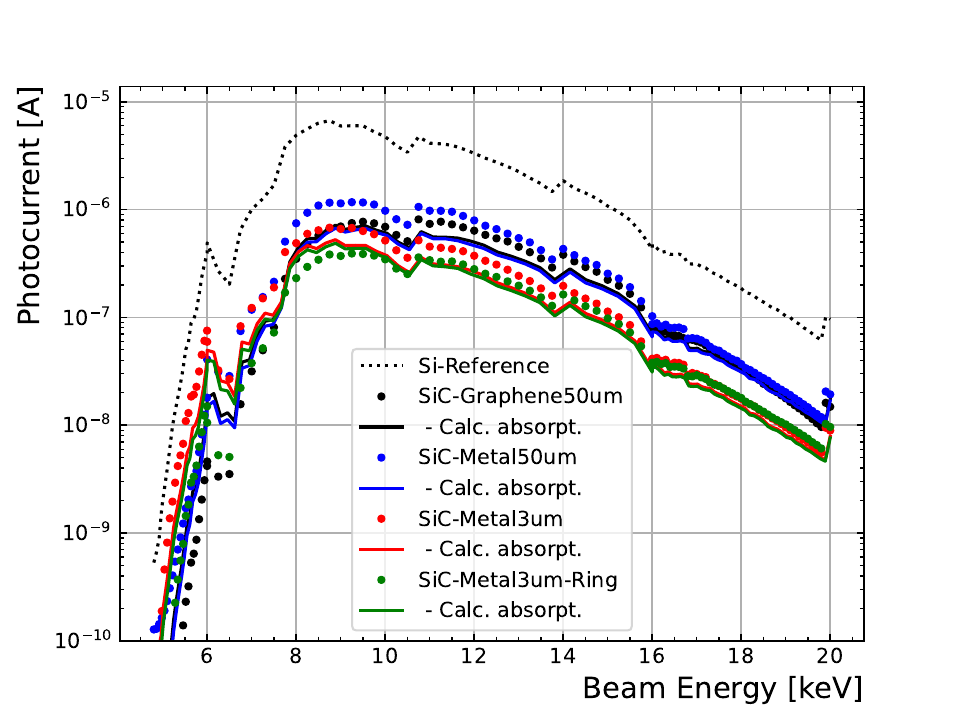}
\caption{Generated photocurrent as a function of beam energy for the Si-Reference detector and the three studied SiC diodes focusing at their respective central region. The same experiment is performed focusing on the bias ring in the 3~$\mu$m epitaxial diode (SiC-Metal3um-Ring). \label{fig:energyscan}}
\end{figure}

As aforementioned, the X-ray beam energy is adjustable between 4.6 and 23~keV range, which allows for an energy response study of the DUTs.
Figure~\ref{fig:energyscan} shows the current measured from each of the SiC diodes when aligning the beam onto the central area of the devices.
Similarly, a calibrated Silicon beam monitor reference detector (Si-Reference) was used to determine the beam power for each chosen X-ray energy.
Finally, in order to assess the effect of the extra metal layers present in the SiC-Metal3um bias ring, the same measurement was performed focusing on that area.

All DUTs show a response consistent with the intensity dependence inherent to the XALOC beam optics reported elsewhere~\cite{XALOC} and observed in the reference device.
Some difference in their respective response are worth mentioning: 
at the highest energies, SiC-Metal3um shows an over-response with respect to its ring due to the effect of the gold, which is reversed at lower energies (<17.5~keV). This is attributed to the metal combination in the ring - which includes the metal stack in the central region, see Figure~\ref{fig:dut-schematic} and Table~\ref{tab:duts}. 
In contrast, the over-response in the SiC-Metal50um with respect to the SiC-Graphene50um is observed over the full energy range.

From the reference detector response, the beam flux ($\phi(E)$), i.e. the number of photons per second in the beam, can be calculated using its calibration. Then, the photocurrent for each device from the photon absorption onto the active area for an energy $E$ is estimated using the following equation:
\begin{equation}
    I(E) = \frac{Eq_{e}\phi(E)}{\varepsilon_{\text{4H-SiC}}}e^{-\sum_{\text{abs}}d_{\text{abs}}/\lambda_{\text{abs}}(E)}\left(1-e^{-d_{\text{SiC}}/\lambda_{\text{SiC}}(E)}\right),
\end{equation}
where $\varepsilon_{\text{4H-SiC}}$=7.8~eV~\cite{SiC-review,SiC-ehpair_1} is the energy needed to create an electron-hole pair in 4H-SiC and $q_{e}$ is the electron charge, while $d$ and $\lambda(E)$ are the thickness and attenuation lengths for the particular energy and absorbing ("abs") layers and the respective SiC active region. 
Table~\ref{tab:simu} compiles the characteristic of the materials used for the calculation and the energy dependent attenuation length are taken from~\cite{xray-trans}.
The result of these calculations (solid lines) agree with the general trend of the experimental results (dots) as shown in Figure~\ref{fig:energyscan}.
Here we use 7~$\mu$m active depths in the SiC-Graphene50um and the SiC-Metal50um as it better reproduces the photocurrent at the highest energies. The small discrepancy with the estimated depletion depth may be a consequence of differences across devices or charge collected by diffusion in the un-depleted volume~\cite{Shi2003}.
Notice however that the photon absorption alone is not able to explain the relative miss-match between the sensors from the measurements, which indicates that the over-response observed is a consequence of secondary electrons from interactions in the metallic contacts~\cite{Metal-dose-enhancement}.

\begin{table}[htbp]
\centering
\caption{Layers considered to calculate the expected photocurrent in each region. After the active area, a bulk SiC of 350~$\mu$m and the corresponding back metal stack are included in the geometry. The numbers refere to those areas marked in Figure~\ref{fig:linescan}. Metal thicknesses are shown in Table~\ref{tab:duts}.\label{tab:simu}}
\smallskip
\begin{tabular}{l|cccc}
\hline\hline
Device           & Cover & Metals& Oxide& Active depth \\ 
\hline
(1) SiC-Graphene50um & 1~mm PMMA & -- & 70~nm Al$_2$O$_3$& 7~$\mu$m SiC    \\
(2)  -Ring          & 1~mm PMMA & Ti/Pt/Au & --& 7~$\mu$m SiC              \\ \hline
(3) SiC-Metal50um    & 1~mm PMMA & Ti/Pt/Au & --& 7~$\mu$m SiC             \\ \hline
(4) SiC-Metal3um     & -- & Ti/Al/Ti/Ni & 800~nm SiO$_2$ & 3~$\mu$m SiC    \\
(5) - Ring         & -- & Ti/Al/Ti/Ni & --& 3~$\mu$m SiC                   \\
                 &     & +Ti/Au       &   &  \\
\hline \hline
\end{tabular}
\end{table}

\section{Conclusions}
\label{sec:conclusions}

Four configurations of silicon carbide radiation detectors fabricated at IMB-CNM have been characterised in the BL13-XALOC beam line at the ALBA Synchrotron.

The presence of some metals such as Au has the effect of increasing the dose reaching the active area of the device when applicable. It introduces an energy dependence which in some cases is detrimental for precision dosimetry.
Amongst others, a 50~$\mu$m epitaxial 4H-SiC diode featuring a graphene layer contact over its active area shows that this effect is significant at low energies. Therefore, the results in this study show that this technology can be utilised for radiation detection with dosimetry purposes.

The 3~$\mu$m epitaxial devices have shown a uniform response consistent to similarly fabricated detectors~\cite{SiC-Graphene_TCT}, even in the case where the current collection is done across the implantation, removing the need of metal in the inner region. However, this may affect its performance after irradiation, as there are indications of loss of dopant concentration, even though further investigation is required.
With a 3~$\mu$m epitaxial geometry, a more consistent behaviour is observed at 0~V bias, owing to the full depletion from the built-in voltage of the pn-junction.

This study shows the characterization of the most recent SiC detector developments at IMB-CNM at low X-ray energies. Results allow us to improve new designs for dosimetry.



\acknowledgments

This work has been funded by the IMB-CNM (CSIC) internal TRIGGER project PlACeD and the laCaixa Health foundation DosiFLASH project HR23-00718.
These experiments were performed at BL13-XALOC beamline at ALBA Synchrotron with the collaboration of ALBA staff. The beam time was funded by ALBA with proposal ID 2024028103.

This work has also received funding from the EMPIR programme co-financed by the Participating States and from the European Union’s Horizon 2020 research and innovation program under project 18HLT04-UHDpulse, from project NEWDOSI (PID2021-123484OB-I00), financed by MCIN / AEI /  10.13039/501100011033 / FEDER, UE and from Grant RTC-2017–6369–3 (GRACE) by MCIN/AEI/ 10.13039/ 501100011033 and ERDF ‘A way of making Europe’.

Support through the Maria de Maeztu grant CEX2023- 001397-M funded by MICIU/AEI/ 10.13039/501100011033 is acknowledged.


\bibliographystyle{JHEP}
\bibliography{biblio.bib}
\end{document}